# Visual Anthropomorphism Shifts Evaluations of Gendered AI Managers

Ruiqing Han[1], Hao Cui[1], and Taha Yasseri[1,2]

[1]Centre for Sociology of Humans and Machines (SOHAM), Trinity College Dublin and Technological University Dublin, Dublin, Ireland
[2]School of Mathematics and Statistics, University College Dublin, Dublin, Ireland

**Abstract**
This research examines whether competence cues can reduce gender bias in evaluations of AI managers and whether these effects depend on how the AI is represented. Across two preregistered experiments ($N = 2,505$), each employing a 2×2×3 design manipulating AI gender, competence, and decision outcome, we compared text-based descriptions of AI managers with visually generated AI faces created using a reverse-correlation paradigm. In the text condition, evaluations were driven by competence rather than gender. When participants received unfavourable decisions, high-competence AI managers were judged as fairer, more competent, and better leaders than low-competence managers, regardless of AI gender. In contrast, when the AI manager was visually represented, competence cues had attenuated influence once facial information was present. Instead, participants showed systematic gender-differentiated responses to AI faces, with feminine-appearing managers evaluated as more competent and more trustworthy than masculine-appearing managers, particularly when delivering favourable outcomes. These gender effects were largely absent when outcomes were unfavourable, suggesting that negative feedback attenuates the influence of both competence information and facial cues. Taken together, these findings show that competence information can mitigate negative reactions to AI managers in text-based interactions, whereas facial anthropomorphism elicits gendered perceptual biases not observed in text-only settings. The results highlight that representational modality plays a critical role in determining when gender stereotypes are activated in evaluations of AI systems and underscore that design choices are consequential for AI governance in evaluative contexts.

## 1. Introduction

As artificial intelligence (AI) systems become increasingly embedded in social and organisational decision-making, understanding how people perceive and evaluate gendered AI has become essential. Evaluations of AI frequently mirror long-standing gender stereotypes: female-coded systems are often perceived as warmer but less competent than male-coded ones (Bastiansen et al., 2022). The present research examines whether explicit competence cues can mitigate gender bias in evaluations of AI managers and whether such effects depend on representational modality, namely text-based versus visually anthropomorphic AI.

### 1.1. Gender Bias in Humans

Gender functions as a primary schema for social categorisation, shaping judgments along the dimensions of agency and communion that underlie perceptions of competence and warmth (Martin & Slepian, 2021). Across cultures, masculinity is more strongly associated with agency and authority, whereas femininity is linked to relational and communal traits (Martin et al., 2024). These associations influence evaluations of competence and leadership and remain resistant to correction because they are largely driven by intuitive, heuristic-based processing rather than deliberative reasoning (Frederick, 2005).

The Stereotype Content Model (SCM) conceptualises social perception along two core dimensions: warmth and competence (Fiske, 2018). Competence, encompassing perceived capability, skill, and agency, is particularly consequential in authority contexts. Decades of research show that women are often expected to be warm but are evaluated more ambivalently on competence, creating tension between being liked and being respected (Connor & Fiske, 2018). These warmth–competence heuristics extend beyond human targets to artificial agents (McKee et al., 2024). In the present research, competence is central because it directly informs judgments of fairness, authority, and leadership in AI-mediated decision-making.

**1.2. Gender, Leadership, and Perceived Competence**

Organisational psychology consistently shows that leadership evaluations depend less on actual performance than on perceived fit between gendered expectations and leadership roles. Meta-analytic evidence indicates that women and men are equally effective leaders overall, yet men are evaluated more favourably in masculine-typed and male-dominated contexts, whereas women receive more positive evaluations in less masculine domains such as education and social services (Eagly et al., 1995). Leadership prototypes remain strongly masculine, particularly in high-status managerial and political settings (Koenig et al., 2011).

Role congruity theory explains these patterns by positing that prejudice toward female leaders arises from a mismatch between descriptive gender stereotypes (women as communal, men as agentic) and the agentic qualities associated with leadership roles (Eagly & Karau, 2002). This perceived incongruity leads women to be judged as less suitable or less competent, even when performance is equivalent, and can also produce backlash when women display agentic behaviour.

The lack-of-fit framework further clarifies how these biases operate (Heilman & Caleo, 2018). When evaluators perceive a discrepancy between stereotypically feminine attributes and the traits believed necessary for success in male-typed roles, they infer lower competence and interpret ambiguous information in ways that disadvantage women. Crucially, the framework predicts that explicit competence cues can reduce bias by diminishing perceived lack of fit.

**1.3. Gendered Perception in Human-AI Interaction**

People readily apply gendered expectations to AI systems, often on the basis of minimal cues (Steeds et al., 2025). A robust male-default bias has been documented, whereby agents are spontaneously categorised as male unless femininity is explicitly signalled (H. Han et al., 2025; Wong & Kim, n.d.). At the same time, AI systems can reflect and reproduce gendered patterns embedded in training data and design choices (Fadaei et al., 2026), meaning that gender is not only projected onto AI but can also be operationalised through its outputs and decision-making behaviour (Yasseri, 2025).

Recent work shows that AI's assigned gender influences human–AI cooperation, with participants exploiting and distrusting agents differently depending on gender labels, mirroring human–human bias patterns (Bazazi, Karpus, & Yasseri, 2025). However, gendered evaluations also depend strongly on task context. Agents performing technical or evaluative functions are often preferred when male-coded, consistent with stereotypes linking competence and authority to masculinity (Jeon, 2024), and female AI managers can receive harsher judgments under unfavourable outcomes compared with their male counterparts (Cui & Yasseri, 2025). Conversely, in relational or caregiving contexts, female agents are often evaluated more positively (Borau et al., 2021).

Crucially, these effects are not uniform. When gender cues are weak or minimally anthropomorphic—such as in text-only chatbot interactions—gender stereotypes often fail to activate (Bastiansen et al., 2022). Gender effects are further moderated by domain and cultural context and can even reverse when stereotypically feminine traits align with valued performance characteristics (e.g., cautious driving behaviour; Pavone & Desveaud, 2025). Overall, prior research suggests that gendered perceptions of AI are selectively triggered by representational cues and situational demands rather than operating uniformly across contexts.

### 1.4. Research Overview

The present research extends recent work on gender bias in AI-mediated leadership by experimentally testing whether competence cues can mitigate the gendered asymmetries observed in evaluations of AI managers. Across two preregistered studies (https://doi.org/10.17605/OSF.IO/TSNGY), we employed parallel designs using textual and visual representations of AI managers to examine whether explicit competence signals influence perceptions of fairness, trust, and leadership suitability following favourable or unfavourable managerial decisions. In Experiment I, competence cues were conveyed through explicit textual descriptions of the AI manager, whereas in Experiment II competence was encoded implicitly through facial representations generated via the reverse-correlation procedure, without any accompanying text. Grounded in Heilman and Caleo's (2018) lack-of-fit model and Eagly and Karau's (2002) role congruity theory, we hypothesised that competence information would reduce the gender bias typically directed at female AI, particularly when its decisions are unfavourable, by enhancing the perceived fit between feminine gender roles and leadership expectations. All the studies received ethics approval (131-LS-LR-25-Yasseri; 132-LS-C-25-Yasseri). All participants provided informed consent prior to participation, and the study complied with institutional and legal ethical standards.

Specifically, we hypothesised that (a) managers who possess high-competence cues would receive more favourable evaluations than low-competence managers when delivering identical unfavourable decisions; (b) this competence advantage would attenuate the harsher judgments typically applied to female, especially female-AI, managers; and (c) in visual conditions, competence cues would alter the mental representations of male and female managerial faces, increasing overlap between high-competence female managers and prototypical leadership imagery. Together, these studies provide an integrated behavioural and perceptual test of how competence cues interact with gender and representation modality to shape human responses to gendered AI authority figures.

Across both experiments, we used an identical task structure and evaluation framework to ensure that any differences in outcomes could be attributed to the representational format of the AI manager rather than to procedural variation. We conducted two experiments to investigate gender bias in perceptions of AI managers. Both experiments employed the same 2×2×3 factorial design, manipulating AI manager competence (high/low), award outcome (selected as best player/not selected), and AI manager gender (female/male/gender-unspecified). The key difference between the experiments was the representation method: in the text-based experiment, AI managers were described in text, whereas in the image-based experiment, AI managers were represented by faces with varying characteristics.

## 2. Experiment I

### 2.1. Method

Experiment I employed the general task structure and experimental design described above. This section details the participants, materials, procedure, and measures specific to the text-based AI manager condition.

#### 2.1.1. Procedure

Prior to the collaborative task, participants completed a pre-survey measuring their initial perceptions of AI managers across dimensions of fairness, leadership, competence, trustworthiness, teammate suitability, and willingness to work with such managers. After completing the collaborative task and receiving the manager's award decision, participants completed a post-survey assessing their evaluations of the manager following direct interaction and outcome feedback (see more details in the supplementary materials S1-S4).

They then participated in a multiplayer online problem-solving game, implemented using the Empirica platform (Almaatouq et al., 2021) and hosted on DigitalOcean servers (https://www.digitalocean.com/), in which they collaborated with other participants to identify a "robber" in a visual puzzle (for more details, see Cui & Yasseri, 2025; also see supplemental materials S10). Participants were informed that an AI manager would evaluate their performance and select the best player to receive an additional monetary reward.

After completing the collaborative task, participants received feedback about whether they were selected as the best player. In reality, this selection was random, although participants were unaware of this during the experiment. Following the manager's decision, participants completed a post-survey assessing the same dimensions as the pre-survey, enabling comparison of perceptions before and after the AI manager's decision.

To manipulate perceived AI manager competence, participants were presented with a brief textual description of the manager prior to receiving the evaluation outcome. The wording was adapted from established agency and competence traits in the leadership literature and varied only in competence level, while holding content length and structure constant across conditions. Managers' descriptions are below (female/male/unspecified):

*High competence:*

"She/He/ The manager demonstrates strong problem-solving abilities, confidently tackles challenges, assertively advocates for her/his ideas against opposition, and takes initiative without guidance."

*Low competence:*

"She/He/ The manager sometimes has difficulty solving complex problems, approaches challenges with caution, tends to give in when her/his ideas face opposition, and prefers guidance before taking on new tasks."

We asked participants to report the manager's gender and whether the manager was human or AI as an attention check. At the conclusion of the experiment, participants were debriefed about the random nature of the selection process.

### 2.1.2. Participants

A total of 1,157 participants were recruited through Prolific. Participants ranged in age from 18 to 79 years ($SD = 13.38$). The sample included 594 males, 556 females, 6 non-binary participants, and 1 participant who identified as "other." Each participant received £1.50 for their participation, and the selected "best player" received an additional £0.50 bonus. A power analysis for between-participants $F$ tests, conducted using the statsmodels package in Python (Seabold & Perktold, 2010; $\alpha = .05$, power = .80), indicated that, with our sample size, the study was sufficiently powered to detect a minimum effect size of $f = 0.121$.

## 2.2. Results

### 2.2.1. Overall Effects of AI Gender, Competence, and Outcome

A series of 3 (AI gender: female, male, unspecified) × 2 (AI competence: high, low) × 2 (evaluation outcome: selected vs. not selected) between-subjects ANOVAs examined perceptions of the AI manager across six dimensions: fairness, trust, competence, leadership, teammate evaluation, and willingness to work with the AI.

Across all dependent variables, evaluation outcome produced the largest and most consistent effects. Participants who were selected as the best player evaluated the AI manager more positively on fairness, $F(1, 1143) = 55.51$, $\eta^2_p = .046$; trust, $F(1, 1143) = 75.95$, $\eta^2_p = .062$; perceived competence, $F(1, 1143) = 78.77$, $\eta^2_p = .064$; leadership, $F(1, 1143) = 80.42$, $\eta^2_p = .066$; teammate evaluation, $F(1, 1143) = 75.02$, $\eta^2_p = .062$; and willingness to work with the AI, $F(1, 1143) = 93.27$, $\eta^2_p = .075$ (all $ps < .001$).

AI competence cues exerted smaller but reliable effects. High-competence AIs were rated as more competent, $F(1, 1143) = 7.11$, $p = .008$, $\eta^2_p = .006$; stronger leaders, $F(1, 1143) = 16.79$, $p < .001$, $\eta^2_p = .014$; better teammates, $F(1, 1143) = 8.62$, $p = .003$, $\eta^2_p = .007$; and more desirable collaborators, $F(1, 1143) = 8.63$, $p = .003$, $\eta^2_p = .007$. Competence did not significantly affect fairness or trust (ps > .05).

Crucially, AI gender had no significant main effects across any dependent variable (all $ps > .10$), and no two- or three-way interactions reached significance. Thus, in the text-based condition, perceptions of AI managers were shaped primarily by outcome valence and, to a lesser extent, competence cues, with no evidence of gender bias.

### *2.2.2. Differential Effects of AI Gender and Competence by Outcome*

To examine whether competence effects varied by outcome valence, separate 2 (Competence) × 3 (Gender) ANOVAs were conducted within each outcome condition (Best Player vs. Not Selected; see Tables 1-2).

When participants were not selected as the best player, AI competence significantly influenced all six outcome measures—fairness, competence, leadership, trust, teammate evaluation, and willingness to work with the AI (all ps ≤ .01; $\eta^2_p = .010–.022$). In contrast, when participants were selected as the best player, competence effects were largely attenuated, emerging only for leadership, $F(1, 372) = 4.12$, $p = .043$, $\eta^2_p = .011$. Across both outcome conditions, AI gender and the Competence × Gender interaction were nonsignificant for nearly all measures.

Change-score analyses further illustrate this asymmetry (Figures 1-2). Positive outcomes increased evaluations across traits, whereas negative outcomes reduced ratings of fairness, competence, and trust. Notably, leadership and willingness to work remained relatively robust to negative feedback when the AI was described as highly competent, suggesting that competence cues buffered adverse reactions under unfavourable conditions.

### 2.3. Discussion

The findings indicate that competence cues matter most when the AI delivers unfavourable outcomes. Under negative feedback, participants relied more heavily on task-relevant competence information: high-competence descriptions buffered adverse reactions, whereas low-competence descriptions amplified them. Although unfavourable decisions reduced perceptions of fairness, trust, and competence overall, emphasising competence helped sustain leadership impressions and willingness to collaborate.

In contrast, favourable outcomes dominated evaluations, rendering competence cues largely redundant. Across all conditions, AI gender and the Competence × Gender interaction were nonsignificant, indicating that gender labels did not systematically bias judgments when presented in a text-only format.

Together, these results suggest that in minimally anthropomorphic contexts, explicit competence information can mitigate negative reactions to AI decisions and preserve engagement, particularly when outcomes are unfavourable.

**Table 1. Two-Way ANOVA Results for Text Condition (Best Player = True)**

| DV | Source | F | p | $\eta^2_p$ |
|---|---|---|---|---|
| Fairness | AI Competence | 0.034 | 0.8545 | 0.000 |
|  | AI Gender | 2.257 | 0.1061 | 0.012 |
|  | Interaction | 1.719 | 0.1807 | 0.009 |
| Competence | AI Competence | 0.714 | 0.3986 | 0.002 |
|  | AI Gender | 1.965 | 0.1416 | 0.010 |
|  | Interaction | 0.899 | 0.4078 | 0.005 |
| Leadership | **AI Competence** | **4.125** | **0.0430** | **0.011** |
|  | AI Gender | 0.866 | 0.4216 | 0.005 |
|  | Interaction | 0.150 | 0.8603 | 0.001 |
| Trust | AI Competence | 0.039 | 0.8438 | 0.000 |
|  | AI Gender | 1.118 | 0.3279 | 0.006 |
|  | Interaction | 0.478 | 0.6205 | 0.003 |
| Teammate | AI Competence | 1.838 | 0.1759 | 0.005 |
|  | AI Gender | 0.212 | 0.8091 | 0.001 |
|  | Interaction | 0.651 | 0.5223 | 0.004 |
| Willingness | AI Competence | 0.692 | 0.4062 | 0.002 |
|  | AI Gender | 1.224 | 0.2951 | 0.006 |
|  | Interaction | 0.167 | 0.8461 | 0.001 |

Note. Table shows *F*, *p*, and partial eta-squared ($\eta^2_p$) for each effect. Values in bold indicate *p* < .05.

**Table 2. Two-Way ANOVA Results for Text Condition (Best Player = False)**

| DV | Source | F | p | $\eta^2_p$ |
|---|---|---|---|---|
| Fairness | **AI Competence** | **8.386** | **0.0039** | **0.011** |
|  | AI Gender | 0.067 | 0.9352 | 0.000 |
|  | Interaction | 1.245 | 0.2886 | 0.003 |
| Competence | **AI Competence** | **10.645** | **0.0012** | **0.014** |
|  | AI Gender | 0.364 | 0.6952 | 0.001 |
|  | Interaction | 0.774 | 0.4615 | 0.002 |
| Leadership | **AI Competence** | **17.509** | **0.0000** | **0.022** |
|  | AI Gender | 0.673 | 0.5103 | 0.002 |
|  | Interaction | 0.031 | 0.9699 | 0.000 |
| Trust | **AI Competence** | **7.665** | **0.0058** | **0.010** |
|  | AI Gender | 0.461 | 0.6310 | 0.001 |
|  | Interaction | 1.088 | 0.3375 | 0.003 |
| Teammate | **AI Competence** | **9.875** | **0.0017** | **0.013** |
|  | AI Gender | 0.920 | 0.3992 | 0.002 |
|  | Interaction | 0.348 | 0.7062 | 0.001 |
| Willingness | **AI Competence** | **14.595** | **0.0001** | **0.018** |
|  | AI Gender | 0.440 | 0.6439 | 0.001 |
|  | Interaction | 0.698 | 0.4978 | 0.002 |

Note. Table shows *F*, *p*, and partial eta-squared ($\eta^2_p$) for each effect. Values in bold indicate *p* < .05.

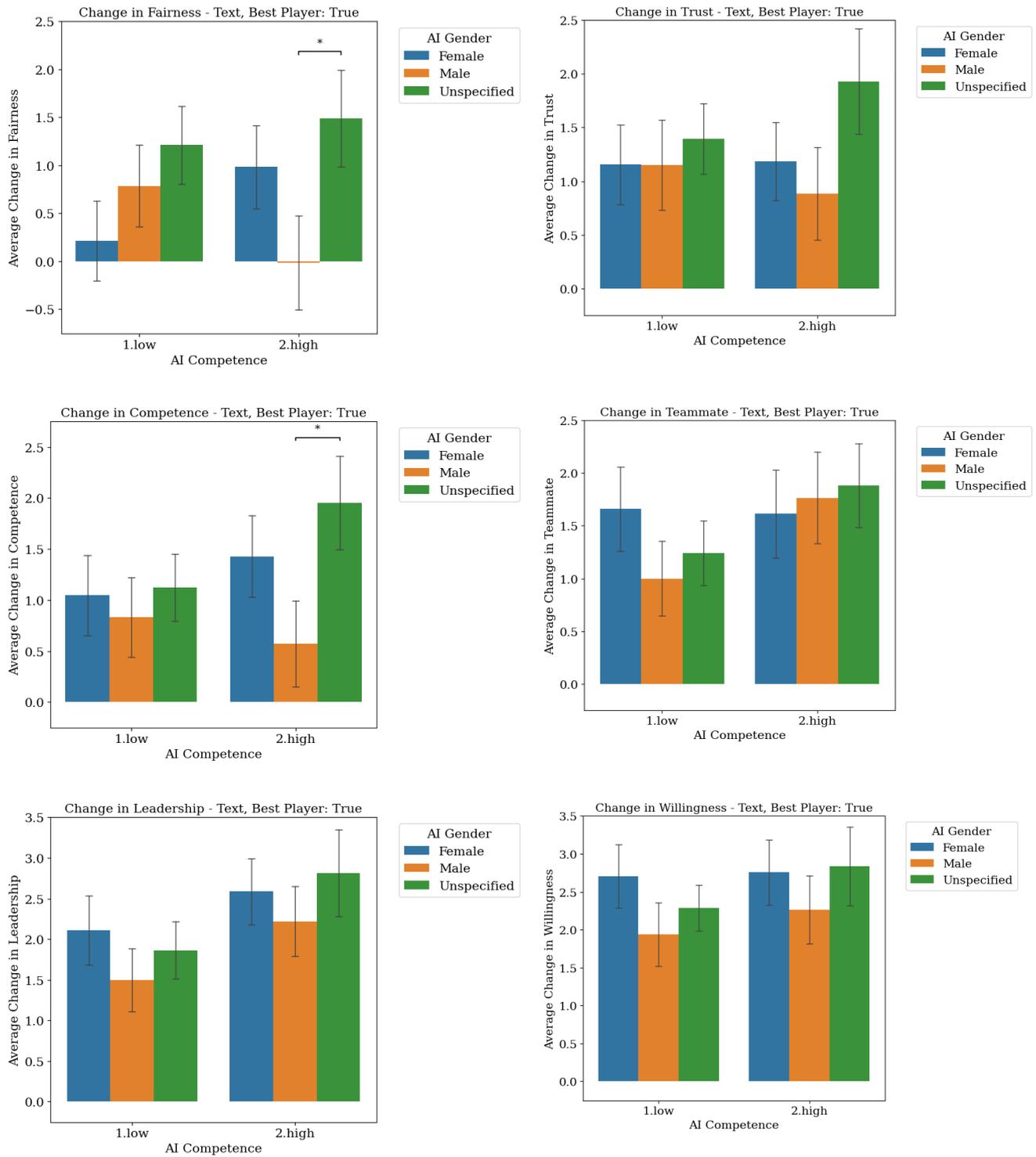

Figure 1. Change in evaluations of the AI manager in the Selected (Best Player) condition as a function of AI competence and AI gender (Experiment I). *Note. Values represent change scores calculated as Δ = Post − Pre survey scores.*

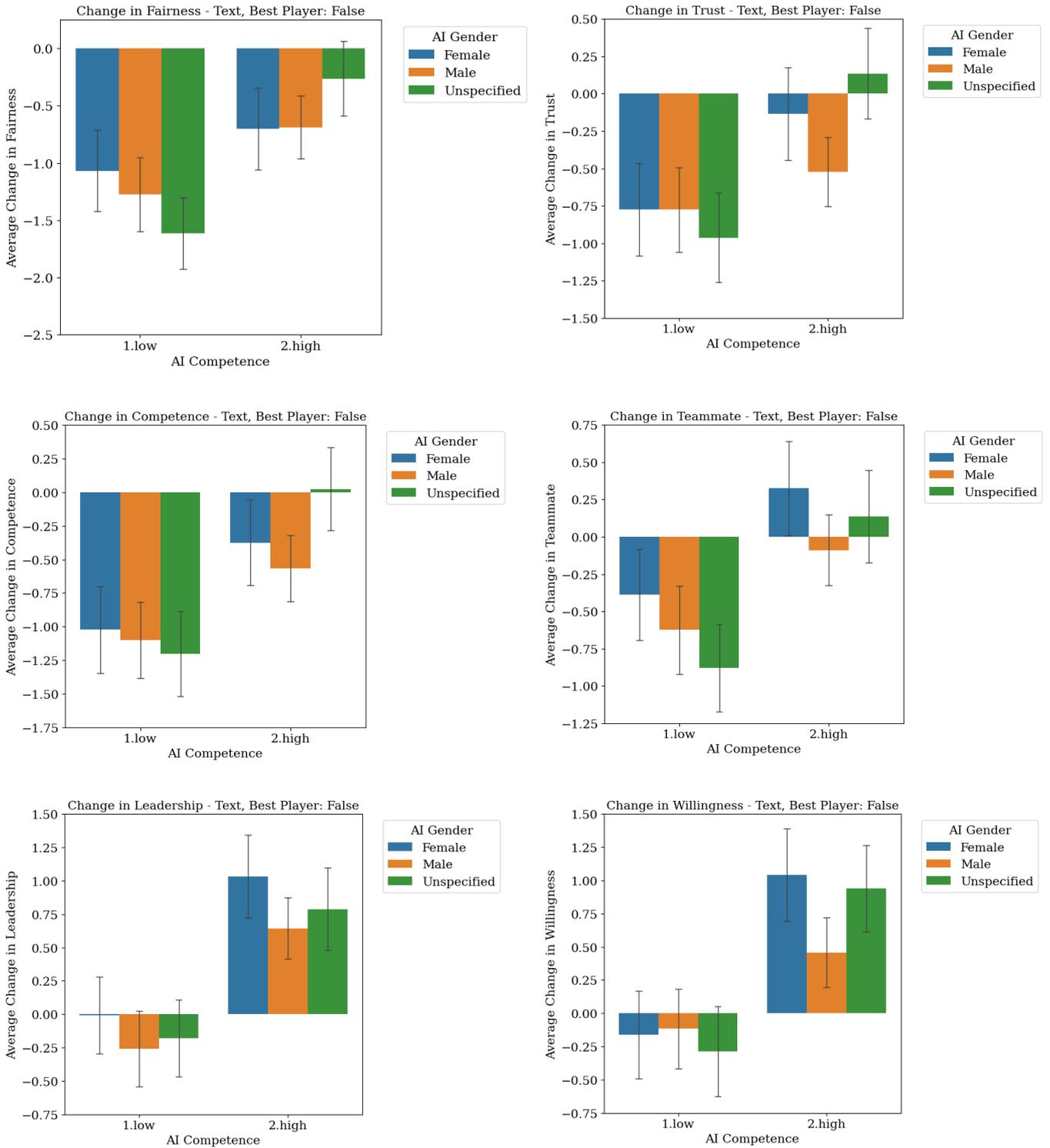

Figure 2. Change in evaluations of the AI manager in the Not Selected condition as a function of AI competence and AI gender (Experiment I). *Note. Values represent change scores calculated as Δ = Post – Pre survey scores.*

## 3. Experiment II

Across human–AI research, anthropomorphism is a central mechanism shaping how people interpret artificial agents. Visual cues, particularly faces, increase social presence and expectations of intentionality, thereby influencing evaluation (Epley et al., 2007; Go & Sundar, 2019; Kim & Im, 2023). However, heightened humanness does not uniformly enhance perceptions; it can also activate social stereotypes and categorical judgments (Troshani et al., 2021). Even gender-ambiguous avatars can elicit gendered interpretations, suggesting that facial cues trigger social categorisation processes that are largely absent in text-based interactions (Aumüller et al., 2024). Together, this work indicates that introducing a face may shift evaluations from deliberative, competence-based judgments toward more intuitive, stereotype-relevant processing.

To examine this shift, Experiment II extended the text-based paradigm by introducing AI manager faces. We employed a reverse-correlation method to generate facial representations that systematically varied in perceived gender and competence. Reverse correlation reconstructs implicit mental prototypes by averaging participants' selections among noise-altered faces (Dotsch & Todorov, 2012), allowing us to capture socially meaningful facial cues without relying solely on explicit self-report. This design enabled direct comparison of whether visual anthropomorphism alters evaluation patterns observed in Experiment I and whether facial gender cues reshape the use of competence information in judgments of AI authority.

### 3.1. Method

Similar to the text-based experiment, participants completed pre- and post-surveys (see more details in the supplementary materials S5-S9) and engaged in the same collaborative problem-solving task. However, instead of text descriptions, participants were shown images of faces representing their assigned AI manager condition. At the end of the experiment, instead of asking participants to identify the AI manager's gender, we asked them to rate how feminine and how masculine they perceived their assigned manager to be. The face images were produced using a computational reverse-correlation paradigm, an algorithmic method that reconstructs facial representations from aggregated visual noise patterns. The procedure is described in detail below.

#### *3.1.1. Reverse Correlation Procedure*

The reverse correlation technique was used to generate face stimuli representing different types of AI managers (Dotsch & Todorov, 2012). The experiment was created using PsychoPy (Pierce et al, 2019) and hosted by Pavlovia (https://pavlovia.org/). Although the main experiment included three gender conditions (female, male, and unspecified) and two competence levels (high and low), the generation phase produced four core gender-specific composites, defined by gender (female or male) and competence (high or low). Generators viewed pairs of noisy images derived from a base face and selected the image that better matched a target trait combination (e.g., "high-competence female" or "low-competence male"). The selected images were then averaged across many trials to create composite faces reflecting the collective perception of each gender-competence combination. Ambiguous-gender faces were created by averaging male and female composites (see Figure 3; for manipulation check of the faces, see Supplementary Materials).

In the second phase, a new sample of participants (raters) rated the generated composite faces on dimensions such as perceived gender, competence, warmth, and other relevant traits. This allowed us to validate whether the generated faces actually conveyed the intended characteristics and to examine additional trait associations that might emerge from these visual representations.

**Generators.** A total of 152 individuals participated in the generating phase. Participants' ages ranged from 18 to 77 years, with a mean age of 42.83 years (*SD* = 13.17). The sample included 76 males, 59 females, and 17 participants who preferred not to disclose their gender. All participants were recruited through Prolific and were compensated £4.50 for their participation.

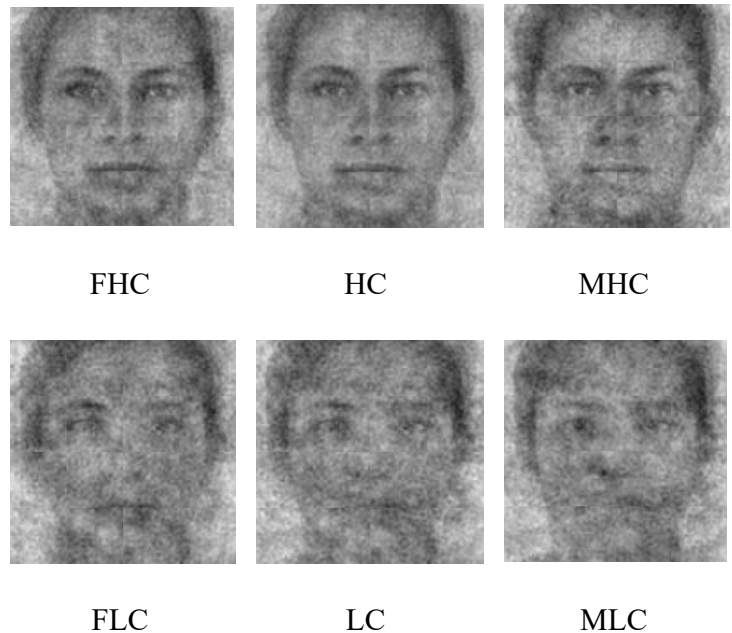

Figure 3. Face stimuli used in Experiment II. Reverse-Correlation Composite Faces for Female (F), Neutral, and Male (M) AI Managers at High-Competence (HC) and Low-Competence (LC) Levels.

**Raters.** A total of 226 individuals participated in the rating phase. Participants' ages ranged from 19 to 77 years, with a mean age of 43.74 years (SD = 13.17). The sample included 122 males, 94 females, and 10 participants who preferred not to disclose their gender. All participants were recruited through Prolific and were compensated £0.50 for their participation.

Independent ratings confirmed that the reverse-correlation procedure successfully produced faces that varied along the intended dimensions. Female composites were rated as more feminine than male composites, and high-competence composites were rated as more competent than low-competence composites across gender conditions. An exploratory factor analysis of interpersonal trait ratings revealed a two-factor structure corresponding to Warmth and Competence, consistent with prior research. Full statistical details, including factor loadings and pairwise comparisons, are reported in the Supplementary Materials.

### 3.1.2. Manager Experiment Procedure

As in the text-based experiment, participants completed pre- and post-surveys measuring their perceptions of AI managers. However, in this version, participants were shown the generated face images representing their assigned AI manager condition. Unlike Experiment I, no textual description of the AI manager's competence and gender was provided; competence and gender cues in Experiment II were conveyed exclusively through the facial representations generated via the reverse-correlation procedure. Participants engaged in the same problem-solving task, received feedback on whether they were selected as the best player (again, randomly determined), and then evaluated their perceptions of the AI manager after experiencing the decision.

*3.1.3. Manager Experiment Participants.*

A total of 1,348 participants took part in the image condition of the study. Participants ranged in age from 18 to 80 years ($SD$ = 13.52). The sample included 613 males, 712 females, 18 non-binary participants, and 4 participants who identified as "other." Each participant was recruited through Prolific and received £1.50 for their participation, with the selected "best player" receiving an additional £0.50 bonus. A sensitivity power analysis for the between-participant F tests, conducted using the statsmodels package in Python (Seabold & Perktold, 2010; $α$ = .05, power = .80), indicated that, with our sample size, the study was sufficiently powered to detect a minimum effect size of $f$ = 0.112.

## 3.2. Results

*3.2.1. Gender Categorisation*

In our reverse-correlation paradigm, we generated three gendered classification images for each condition: male, female, and neutral versions. Although these images were created to represent distinct gender categories, we also collected participants' perceived femininity and masculinity ratings in the main experiment (see Figure 4). These ratings revealed substantial individual variability. Some participants judged the "male" image as relatively feminine, some judged the "female" image as relatively masculine, and the "neutral" image was variously perceived as male, female, or neutral (i.e., equally masculine and feminine).

This variability is expected: reverse-correlation images differ in gender appearance on average, but individual observers can still interpret the same face differently. For this reason, rather than relying on the manipulated gender labels, we base our analyses on participants' perceived gender ratings, which provide a more accurate, participant-specific measure of how gendered each image appears.

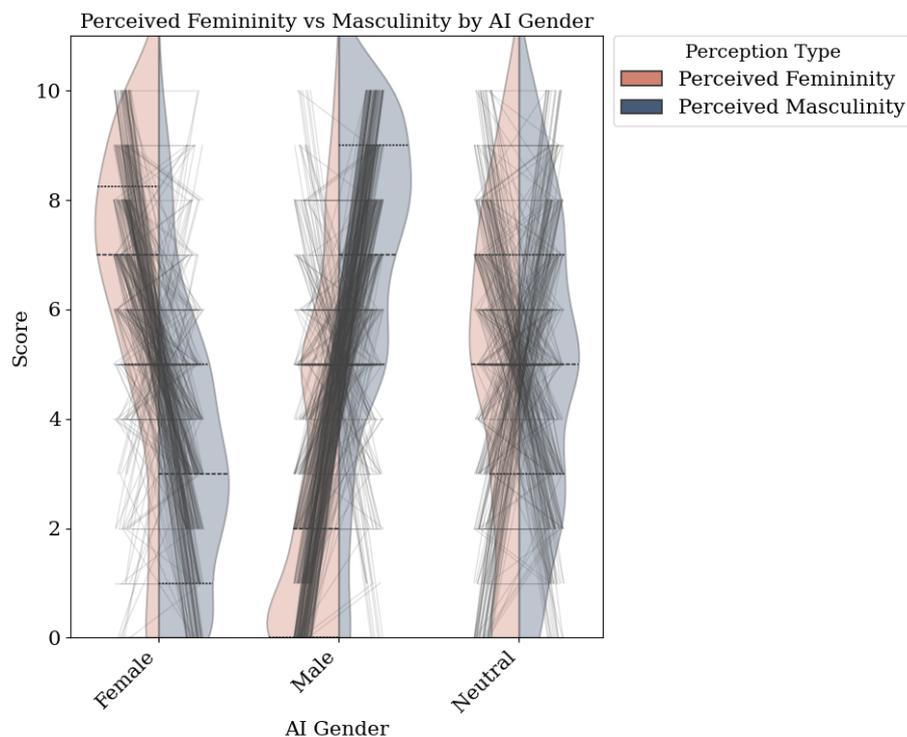

Figure 4. Perceived gender ratings (femininity and masculinity) of AI faces across manipulated gender conditions. Violin plots show the distribution of perceived femininity and masculinity ratings by AI gender condition. Thin connecting lines represent within-participant perceptions.

We conducted a principal component analysis (PCA) on participants' femininity and masculinity ratings to examine how these two judgments jointly contribute to perceived gender. The first principal component accounted for 76% of the variance and reflected a single underlying dimension of gender perception. Femininity and masculinity loaded on this component with equal magnitude but opposite signs (loadings = −0.71 and 0.71), indicating that they represent opposite ends of the same perceptual continuum. Thus, femininity and masculinity contributed equally to participants' gender perceptions.

Based on this, we defined gender categorisation in our main analysis using the following rule: Faces were categorised as feminine when femininity ratings exceeded masculinity ratings, masculine when masculinity exceeded femininity, and ambiguous when ratings were equal.

### 3.2.2. Overall Effects of AI Gender, Competence, and Outcome

A series of 3 (perceived AI gender: feminine, masculine, ambiguous) × 2 (AI competence: high vs. low) × 2 (evaluation outcome: selected vs. not selected) ANOVAs examined how facial anthropomorphism shaped evaluative change.

Consistent with Experiment I, outcome valence remained the strongest determinant across all dependent variables: fairness, $F(1, 1333) = 103.73$, $\eta^2_p = .072$; trust, $F(1, 1333) = 103.10$, $\eta^2_p = .072$; competence, $F(1, 1333) = 82.93$, $\eta^2_p = .059$; leadership, $F(1, 1333) = 121.83$, $\eta^2_p = .084$; teammate evaluation, $F(1, 1333) = 102.21$, $\eta^2_p = .071$; and willingness to work with the AI, $F(1, 1333) = 107.88$, $\eta^2_p = .075$ (all $p$s < .001).

Unlike the text-based condition, perceived AI gender produced reliable differences for trust, $F(2, 1333) = 4.16$, $p = .016$, $\eta^2_p = .006$, and perceived competence, $F(2, 1333) = 6.23$, $p = .002$, $\eta^2_p = .009$. Pairwise comparisons indicated that feminine faces were rated as more trustworthy and more competent than masculine faces, while ambiguous faces fell between conditions without significant differences. No gender effects emerged for fairness, leadership, teammate evaluation, or willingness to work.

In contrast, AI competence cues exerted minimal influence. A small main effect emerged for perceived competence, $F(1, 1333) = 4.22$, $p = .040$, $\eta^2_p = .003$, but competence did not significantly affect other outcomes. No two- or three-way interactions reached significance.

Overall, Experiment II revealed a qualitative shift relative to the text-based condition. While outcome valence continued to dominate evaluations, the addition of facial cues introduced gender-differentiated judgments absent in Experiment I. Specifically, masculine-looking AI managers received lower trust and competence ratings than feminine-looking managers, whereas competence information embedded in facial cues had little impact. These findings suggest that visual anthropomorphism shifted evaluation toward gendered heuristics and reduced the influence of competence cues.

### 3.2.3. Differential Effects of AI Gender and Competence by Outcome

To probe outcome-specific effects, separate 2 (Competence) × 3 (Perceived Gender) ANOVAs were conducted within each outcome condition (see Tables 3-4).
When participants were selected as the best player, perceived AI gender produced reliable differences in several domains. Feminine faces were evaluated more positively than masculine faces on competence, $F(2, 431) = 4.46$, $p = .012$, $\eta^2_p = .020$, and trust, $F(2, 431) = 4.56$, $p = .011$, $\eta^2_p = .021$, with a marginal effect for fairness, $F(2, 431) = 3.00$, $p = .051$, $\eta^2_p = .014$. AI competence also influenced perceived competence ratings, $F(1, 431) = 4.18$, $p = .042$, $\eta^2_p = .009$. No significant interactions emerged, and leadership, teammate evaluation, and willingness showed no overall main effects at this level.

In contrast, when participants were not selected as the best player, neither perceived gender nor competence significantly affected any outcome measure (all $p$s > .22), and no interactions were observed. Negative feedback uniformly attenuated evaluations regardless of the AI's gender or competence cues.

Pairwise comparisons (Figures 5-6) provided additional insight into how perceived AI gender shaped evaluations within each competence condition. In the unfavourable-outcome condition, differences were minimal; the only significant effect was observed in the high-competence condition, in which feminine AI managers were perceived as more suitable for leadership than the ambiguous-gender manager, and more competent than the masculine manager. No other pairwise differences were observed in this outcome condition, suggesting that negative feedback largely overshadowed facial gender cues. By contrast, when participants were selected as the best player, gender differences emerged only in the high-competence condition. In this context, feminine AI managers were perceived as fairer than ambiguous managers, and as more competent and more suitable for leadership than masculine managers. They were also rated as more trustworthy, more desirable as teammates, and more willing to work with than masculine managers. No gender differences appeared in the low-competence condition. Thus, favourable outcomes amplified the impact of facial gender cues, but only when the AI was highly competent, yielding a consistent advantage for feminine-appearing managers over both masculine and ambiguous faces across multiple social and performance-relevant traits.

Overall, whereas competence buffered negative reactions in the text-based experiment, the image condition showed a different pattern: facial gender cues influenced evaluations primarily under favourable outcomes and particularly when competence was high, producing a consistent advantage for feminine-looking AI managers.

### 3.3. Discussion

Experiment II examined whether anthropomorphic facial cues elicit gendered evaluation patterns when users interact with visually represented AI managers. In contrast to the text-based experiment, in which competence cues shaped evaluations, and AI gender had no effect, facial representations elicited reliable gender-differentiated responses. Although outcome valence again dominated impressions, participants consistently rated feminine-looking AI managers as more trustworthy and more competent than masculine-looking AI managers, particularly when the AI delivered favourable outcomes. These gender asymmetries did not emerge in the low-competence conditions and were largely absent when the AI delivered unfavourable decisions, suggesting that negative feedback overshadowed both gender and competence cues.

The presence of facial representations appears to shift evaluation toward rapid person-perception processes, reducing the weight of task-relevant competence information. In this context, masculine-looking AI managers were systematically evaluated less favourably across several dimensions, indicating that facial cues activated gendered heuristics that reshaped the evaluation patterns observed in the textual condition.

Overall, the introduction of human-like faces fundamentally altered how AI managers were judged. Rather than stabilising perceptions, anthropomorphic cues redirected attention toward social categorisation processes, allowing gender-based inferences to guide judgments of trust, competence, and authority.

**Table 3. Two-Way ANOVA Results for Image Condition (Best Player = True)**

| DV | Source | F | p | $\eta^2_p$ |
|---|---|---|---|---|
| **Fairness** | AI Competence | 0.082 | 0.7743 | 0.000 |
| | **AI Gender** | **3.003** | **0.0507** | **0.014** |
| | Interaction | 0.981 | 0.3757 | 0.005 |
| **Competence** | **AI Competence** | **4.178** | **0.0416** | **0.009** |
| | **AI Gender** | **4.461** | **0.0121** | **0.020** |
| | Interaction | 0.002 | 0.9983 | 0.000 |
| **Leadership** | AI Competence | 1.334 | 0.2487 | 0.003 |
| | AI Gender | 1.662 | 0.1910 | 0.008 |
| | Interaction | 2.043 | 0.1308 | 0.009 |
| **Trust** | AI Competence | 0.446 | 0.5046 | 0.001 |
| | **AI Gender** | **4.555** | **0.0110** | **0.021** |
| | Interaction | 0.113 | 0.8931 | 0.001 |
| **Teammate** | AI Competence | 2.061 | 0.1518 | 0.005 |
| | AI Gender | 0.680 | 0.5070 | 0.003 |
| | Interaction | 1.539 | 0.2157 | 0.007 |
| **Willingness** | AI Competence | 0.066 | 0.7969 | 0.000 |
| | AI Gender | 1.642 | 0.1948 | 0.008 |
| | Interaction | 2.106 | 0.1229 | 0.010 |

*Note.* Table shows *F*, *p*, and partial eta-squared ($\eta^2_p$). Bold values indicate $p \leq .05$.

**Table 4. Two-Way ANOVA Results for Image Condition (Best Player = False)**

| DV | Source | F | p | $\eta^2_p$ |
|---|---|---|---|---|
| **Fairness** | AI Competence | 0.374 | 0.5407 | 0.000 |
| | AI Gender | 0.014 | 0.9858 | 0.000 |
| | Interaction | 0.933 | 0.3937 | 0.002 |
| **Competence** | AI Competence | 1.487 | 0.2231 | 0.002 |
| | AI Gender | 2.901 | 0.0555 | 0.006 |
| | Interaction | 0.256 | 0.7741 | 0.001 |
| **Leadership** | AI Competence | 0.747 | 0.3877 | 0.001 |
| | AI Gender | 0.963 | 0.3821 | 0.002 |
| | Interaction | 1.159 | 0.3144 | 0.003 |
| **Trust** | AI Competence | 0.089 | 0.7660 | 0.000 |
| | AI Gender | 0.498 | 0.6081 | 0.001 |
| | Interaction | 0.800 | 0.4496 | 0.002 |
| **Teammate** | AI Competence | 0.024 | 0.8758 | 0.000 |
| | AI Gender | 0.090 | 0.9136 | 0.000 |
| | Interaction | 1.579 | 0.2068 | 0.004 |
| **Willingness** | AI Competence | 0.318 | 0.5729 | 0.000 |
| | AI Gender | 0.418 | 0.6586 | 0.001 |
| | Interaction | 0.867 | 0.4204 | 0.002 |

*Note.* Table shows *F*, *p*, and partial eta-squared ($\eta^2_p$). No effects reached significance (all *p*s > .05).

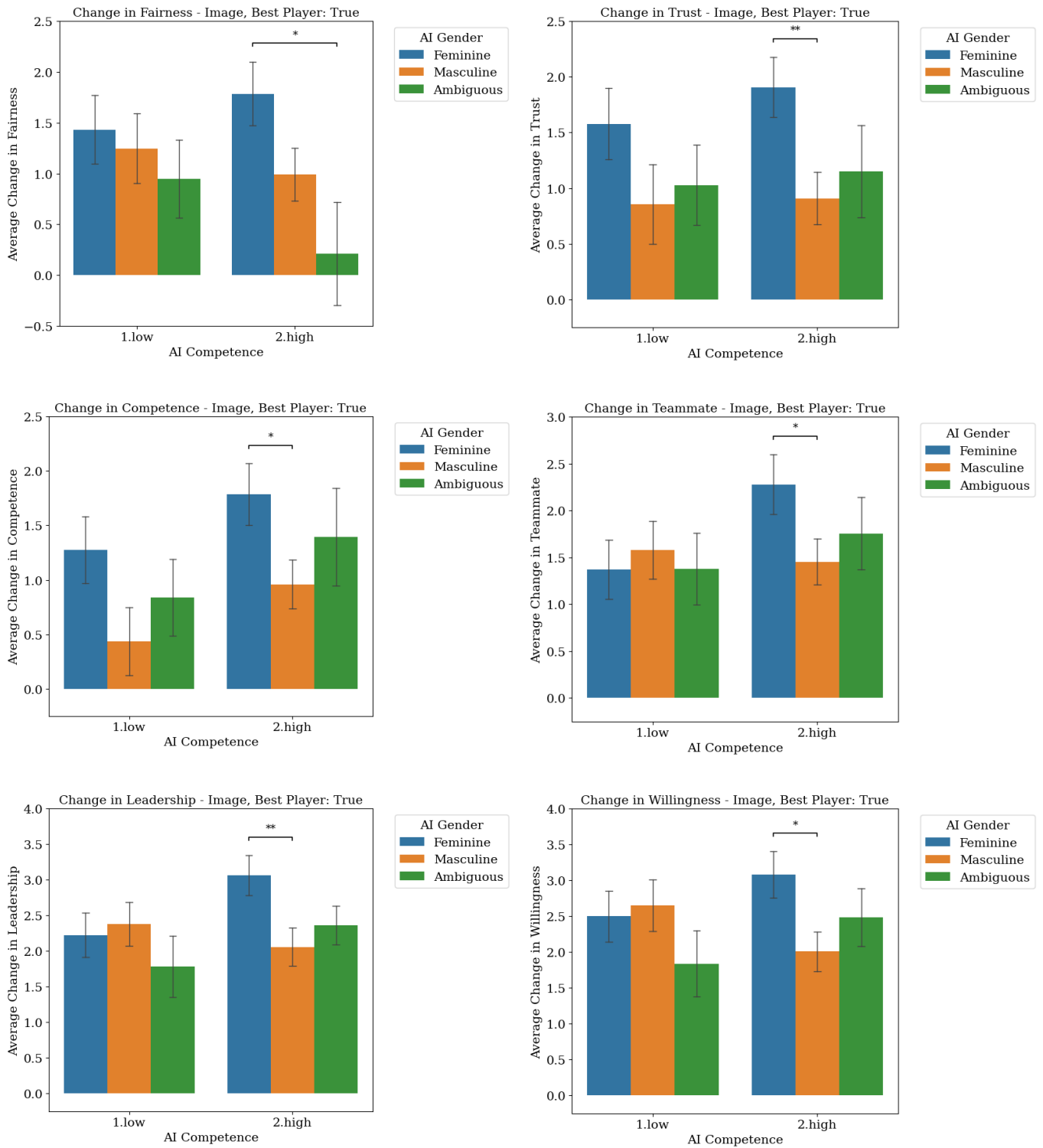

Figure 5. Change in evaluations of the AI manager in the Selected (Best Player) condition as a function of AI competence and AI gender (Experiment II). *Note. Values represent change scores calculated as Δ = Post − Pre survey scores.*

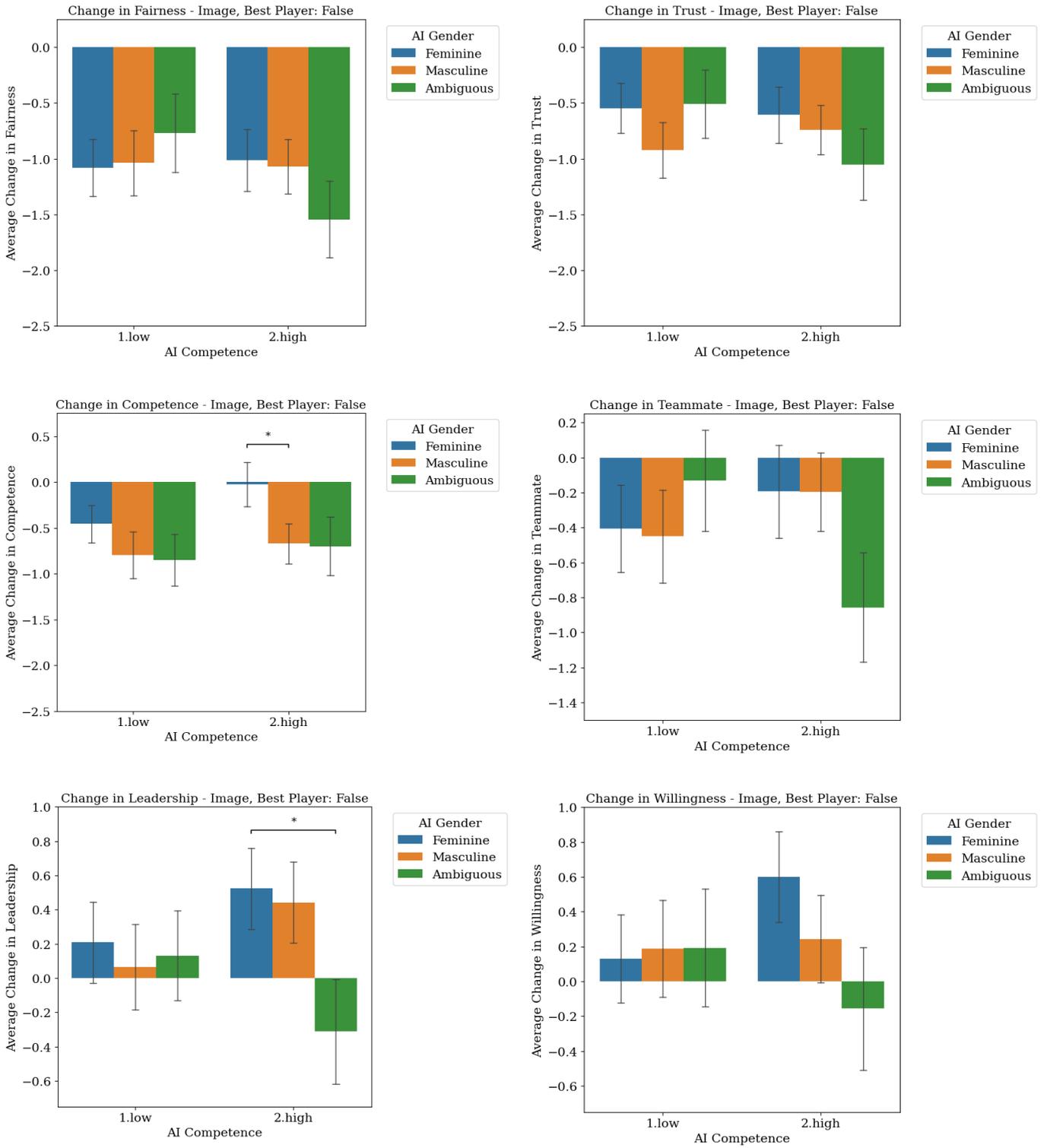

Figure 6. Change in evaluations of the AI manager in the Not Selected condition as a function of AI competence and AI gender (Experiment II). *Note. Values represent change scores calculated as Δ = Post – Pre survey scores.*

# 4. General Discussion

The present research examined whether competence cues can mitigate gender bias in evaluations of AI managers and whether these effects depend on how the AI is represented. Across two large-scale experiments using parallel designs, we found that competence signals meaningfully shaped evaluations of text-based AI managers but were substantially attenuated once the AI was rendered visually. These findings clarify how gendered expectations are selectively activated, suppressed, or redirected in human-AI interaction.

## 4.1. Summary of Findings

In the text-based condition, competence served a consistent buffering function. Under unfavourable feedback, high-competence AI managers were evaluated more positively across leadership, trust, and willingness to collaborate, whereas low-competence managers were penalised more strongly. AI gender had no measurable effect. These findings are consistent with evidence that gender cues often fail to activate in minimally anthropomorphic environments and that competence plays a central role in trust formation in AI systems (Bastiansen et al., 2022; Christoforakos et al., 2021).

In contrast, introducing facial anthropomorphism produced a qualitatively different pattern. Once faces were present, gendered evaluations emerged across several dimensions despite the same underlying task structure. Masculine-looking AI managers were trusted less and perceived as less competent than feminine-looking managers, particularly when delivering favourable outcomes and in high-competence conditions. This pattern aligns with research showing that human-like embodiments increase the salience of social stereotypes (Bernotat et al., 2021; Eyssel & Hegel, 2012). Notably, negative outcomes largely attenuated these differences, suggesting that unfavourable decisions dampen the influence of both competence cues and facial gender.

Taken together, the findings indicate that competence cues guide evaluation in text-only contexts, whereas facial anthropomorphism activates gendered heuristics that reshape how identical performance information is interpreted. Although our preregistered hypothesis focused on bias directed at female AI managers, the observed gender asymmetries in the visual condition instead manifested as a relative disadvantage for masculine-looking AI managers under favourable outcomes.

## 4.2. Competence as a Buffer and Its Limits

The divergence across modalities highlights the conditional influence of competence cues. In text-based interactions, competence information was salient and diagnostic, guiding evaluations and buffering negative reactions, consistent with prior work emphasising competence in trust formation under low social-information conditions (Christoforakos et al., 2021). Once facial cues were introduced, however, competence no longer structured judgments to the same extent.

In the image condition, competence signals were perceptually embedded in facial representations, yet they exerted only limited influence on evaluation. This pattern suggests not an absence of competence information but a shift in processing: the introduction of faces redirected attention toward social categorisation mechanisms inherent to person perception. When faces resemble socially meaningful categories, these processes readily activate stereotype-relevant inferences (Bernotat et al., 2021; Eyssel & Hegel, 2012), thereby diminishing the relative weight of task-relevant competence information.

The pattern observed here is consistent with the Stereotype Content Model (Fiske, 2018). Text-based AI managers elicited evaluations primarily along a competence dimension, whereas facial representations activated both competence and warmth-related inferences, increasing the salience of gendered expectations. Feminine-looking AI managers were consequently perceived as more trustworthy and competent under favourable conditions, while negative outcomes attenuated these differences.

Taken together, the findings indicate that representational modality determines whether competence cues remain dominant or are subsumed by gendered social heuristics triggered by anthropomorphic facial cues.

### 4.3. Implications for Gendered AI Design

These findings have important implications for AI design. Prior work shows that algorithmic systems already reproduce gender bias through training data and design choices (Hall & Ellis, 2023). Our results demonstrate that visual representation alone, even when competence is controlled, can introduce new forms of bias that do not arise in text-only contexts. Anthropomorphism therefore appears to function as an amplifying mechanism for stereotypes rather than a neutral design choice.

At the same time, evidence suggests that design interventions can mitigate bias. Allowing users to select an AI's gender can prompt stereotype-correcting behaviour, particularly among women choosing counter-stereotypical agents in male-dominated domains (Claudy et al., 2025). Together with our findings, this suggests that gender-neutral or user-selected representations may reduce bias by preventing automatic gender activation from visual cues. Moreover, the asymmetry observed in our image-based results parallels broader patterns in AI gendering. Female agents are often deployed in warmth-oriented roles and male agents in evaluative or technical roles (Borau et al., 2021). Our findings challenge the assumption that male representations necessarily confer advantages in competence-related contexts, particularly when users must interpret reward or evaluation decisions.

### 4.4. Broader Theoretical Contributions

This research contributes to the literature on gender and AI by demonstrating that gender bias in AI evaluation is highly sensitive to representational modality. In text-based interactions, competence cues reliably shape evaluations, whereas visually represented AI activates gendered perceptual heuristics that diminish the influence of task-relevant competence information. Competence and gender cues thus play distinct roles across modalities, with perceived gender becoming a stronger predictor of trust formation than competence descriptions in the image condition.

These findings extend prior work on competence and anthropomorphism (Christoforakos et al., 2021) by showing that the salience of gendered facial cues determines whether competence information is incorporated into evaluations of AI managers. They also provide empirical evidence that masculine anthropomorphic AI may be disadvantaged in accountability-related evaluative contexts, extending research on anthropomorphism, perceived intentionality, and emotional reactions to AI (Garvey et al., 2023), although this disadvantage emerged primarily under favourable outcomes.

### 4.5. Limitations

Several limitations should be noted. First, the reverse-correlation faces were not equally noisy across competence conditions and did not always cleanly reflect manipulated gender categories. This is an expected feature of reverse-correlation methods: high-competence categories tend to yield more internally consistent selections and smoother composites, whereas low-competence categories produce noisier images. Similarly, variability in how participants interpreted "male" and "female" composites is consistent with prior work showing individual differences in such representations (Albohn et al., 2025). Although this reduces experimental control, it captures participants' own gender perceptions, which is ecologically relevant for understanding anthropomorphic AI evaluation.

Second, the "unspecified" gender condition differed across modalities: absence of a cue in text versus a blended face in images. While these operationalisations are not identical, the discrepancy mirrors real-world gender inference, where visual cues often trigger gender categorisation even without labels. Third, outcome valence strongly shaped evaluations in both studies, potentially muting subtler competence or gender effects, particularly under unfavourable outcomes. Future work using repeated interactions or more gradual feedback may reveal dynamics not captured by single-outcome designs.

Taken together, these limitations highlight areas for future work to refine stimulus control, examine repeated interactions, or further disentangle facial cues. At the same time, they also underscore that many of the ambiguities we encountered, especially around gender perception, are not merely methodological artefacts but core features of how people actually perceive and evaluate AI systems.

## 5. Conclusion

Taken together, these findings show that task-relevant competence cues can mitigate negative reactions in text-based interactions, but visual anthropomorphism introduces new patterns of bias, creating relative disadvantages for masculine-looking AI managers under favourable outcomes. These results highlight the importance of representational design in AI systems. Designers and policymakers should consider the unintended consequences of anthropomorphic features, particularly facial ones, and recognise that gender-neutral or user-selected representations may help minimise bias. As the field advances, integrating technical audits of algorithmic outputs with attention to social perception processes will be essential for developing AI systems that are both effective and equitable.

**Acknowledgements**

We thank Faezeh Fadaei for her technical support. We also thank Jenny Moran and Marius Claudy for sharing their insights on the research.

**Funding Statement**

The research conducted in this publication was funded by Research Ireland under grant number IRCLA/2022/3217, ANNETTE (Artificial Intelligence Enhanced Collective Intelligence). TY thanks Workday Inc for support.

**Data Statement**

Anonymised data of this study are available at https://osf.io/6m8hs/files/zwjcf.

**Author Contributions (CRediT)**

Ruiqing Han: Conceptualisation, Software, Methodology, Investigation, Data curation, Formal analysis, Visualisation, Writing - original draft, Writing - review & editing.

Hao Cui: Software, Methodology, Writing - review & editing.

Taha Yasseri: Conceptualisation, Methodology, Supervision, Funding acquisition, Writing - review & editing.

**Generative AI Disclosure**

During the preparation of this manuscript, the authors used generative AI tools for language editing to improve clarity and grammar. The authors reviewed and edited the content and take full responsibility for the final manuscript.

# Supplementary Materials

**Survey Questions for Experiment I (Text-Based AI)**

**S1. Pre-Survey Measures (Baseline)**

Before the collaborative task, participants completed baseline perception surveys. All items were presented as single-item measures using 11-point response scales (0–10) implemented as radio-button responses, and all items were mandatory (participants could not proceed until all required items were answered).

*S1.1 AI Manager Evaluation (Baseline)*

Participants rated a trained "AI" manager on six dimensions:
1. "How trustworthy do you think a trained 'AI' manager is in making award decisions for the team they manage?
   0: very untrustworthy; 10: very trustworthy"
2. "How competent do you think a trained 'AI' manager is in making award decisions for the team they manage?
   0: very incompetent; 10: very competent"
3. "How fair do you think a trained 'AI' manager is in making award decisions for the team they manage?
   0: very unfair; 10: very fair"
4. "Would you be willing to work in a small team led by a trained 'AI' manager who makes award decisions for the team they manage?
   0: very unwilling; 10: very willing"
5. "How suitable do you think a trained 'AI' manager is for leadership roles?
   0: very unsuitable; 10: very suitable"
6. "How suitable do you think a trained 'AI' would be as a teammate?
   0: very unsuitable; 10: very suitable"

*S1.2 Human Manager Evaluation*

Participants also rated an experienced "human" manager on the same six dimensions using parallel wording (trustworthiness, competence, fairness, willingness, leadership suitability, and teammate suitability), with the manager described as an experienced "human" manager.

*S1.3 Male and Female Manager Evaluation*

Participants rated an experienced "male" manager and an experienced "female" manager on the same six dimensions, using parallel items that explicitly referenced manager gender (e.g., "experienced 'male' manager" and "experienced 'female' manager").

**S2. Pre-Survey Order Randomisation**

Two counterbalancing procedures were implemented and stored per participant:
1. **Male vs. Female manager pages**: participants saw either the male page first or the female page first (randomised).
2. **Human vs. AI manager pages**: participants saw either the human page first or the AI page first (randomised).

**S3. Post-Survey Measures (Text Experiment)**

After the award decision, participants completed a post-survey.

*S3.1 Satisfaction*

Participants rated satisfaction with the award decision:

- "How satisfied did you feel about the award decision based on the teammate evaluations in the game?
  0: very unsatisfied; 10: very satisfied"

*S3.2 Manager Identification (Attention Checks)*

Participants identified the manager type and manager gender:

1. "What was the manager type in your group?"
   Options: Human, AI, No manager
2. "What was the manager gender in your group? (Please select 'Gender unspecified' if no manager was assigned.)"
   Options: Female, Male, Gender unspecified

*S3.3 Open-Ended Comments*

Participants could provide written feedback:
- "Please provide any additional comments or insights regarding your experience with the experiment."

**S4. Exit Survey (Text Experiment)**

After the collaborative task, participants completed an additional exit survey evaluating the manager's performance and suitability. All items used 11-point scales (0–10) and were mandatory.

1. "How trustworthy do you think this manager's evaluation was in making award decision for the best player in the game?
   0: very untrustworthy; 10: very trustworthy"
2. "How competent do you think this manager's evaluation was in making award decision for the best player in the game?
   0: very incompetent; 10: very competent"
3. "How fair do you think this manager's evaluation was in making award decision for the best player in the game?
   0: very unfair; 10: very fair"
4. "Will you be willing to work in future teams like the one in this game, where award decisions are determined based on this manager's evaluations?
   0: very unwilling; 10: very willing"
5. "How suitable do you think this manager is for leadership roles?
   0: very unsuitable; 10: very suitable"
6. "How suitable do you think this manager would be as a teammate?
   0: very unsuitable; 10: very suitable"

**Survey Questions for Experiment II (Image-Based AI)**

**S5. Pre-Survey Measures (Baseline)**

The pre-survey measures and order randomisation procedures in Experiment II were identical to Experiment I (AI manager baseline evaluation, human manager baseline evaluation, male and female manager baseline evaluations, and counterbalancing).

**S6. Post-Survey Measures (Image Experiment)**

After the award decision, participants completed a post-survey containing satisfaction and manager-type identification items. All required items were mandatory.

*S6.1 Satisfaction*
- "How satisfied did you feel about the award decision based on the teammate evaluations in the game?
  0: very unsatisfied; 10: very satisfied"

*S6.2 Manager Type Identification*
- "What was the manager type in your group?"
Options: Human, AI, No manager

*S6.3 Open-Ended Comments*
Participants could provide written feedback:
- "Please provide any additional comments or insights regarding your experience with the experiment."

**S7. Exit Survey (Image Experiment)**

At the end of the image experiment, participants completed an exit survey while being shown a visual reference image (visual.png) on screen. All items used 11-point scales (0–10) and were mandatory.

**Perceived gender appearance items (presented first):**
1. "How feminine do you think this manager appeared?
0: very low; 10: very high"
2. "How masculine do you think this manager appeared?
0: very low; 10: very high"

**Manager evaluation items (presented after the gender appearance items):**
3. "How competent do you think this manager appears?
0: very low; 10: very high"
4. "How trustworthy do you think this manager's evaluation was?
0: very low; 10: very high"
5. "How fair do you think this manager's evaluation was?
0: very low; 10: very high"
6. "Will you be willing to work in future teams like this?
0: very low; 10: very high"
7. "How suitable do you think this manager is for leadership?
0: very low; 10: very high"
8. "How suitable do you think this manager would be as a teammate?
0: very low; 10: very high"

**S8. Demographics Survey**

All participants completed a demographics survey at the end of the study. Items were mandatory.

1. **Age** (numeric entry; minimum 18, maximum 100):
"Age"
2. **Gender identity** (single choice):
"What is your gender identity?"
Options: Male, Female, Non-binary, Other
3. **Education** (single choice):
"What is the highest level of education you have completed?"
Options:
- Secondary education
- Undergraduate degree (e.g., BA, BSc)
- Postgraduate degree (e.g., MA, MSc, PhD)
- Other

## S9. Response Validation

Across surveys in both experiments, required questions enforced response completion prior to continuing. For radio-based items, participants could not proceed unless a value was selected. For demographics, age, gender identity, and education were required.

## S10. Task Procedure and Interface Flow (Experiments 1 and 2)

This section provides a detailed, step-by-step description of the experimental task (see Cui & Yasseri), interface flow, and participant experience common to both Experiment I (Text-Based AI Manager) and Experiment II (Image-Based AI Manager). The two experiments were intentionally designed to be procedurally identical, differing only in how the AI manager was represented (textual description vs. facial image). Below, we first describe the shared task structure in full (S10.1–S10.8), corresponding to the screenshots shown in Supplementary Figures S1–S4. We then specify the representation-specific differences for Experiment I and Experiment II (S10.9–S10.10).

### S10.1 Overview

Both experiments employed the same interactive, multi-stage online task intended to simulate a collaborative work setting in which an AI manager evaluates team performance and allocates a reward. Participants completed the study online and interacted in real time with two other human participants. The task consisted of two rounds: an individual problem-solving round and a team-based round followed by managerial evaluation.

### S10.2 Round 1: Individual Task

Participants first completed an individual round in which they were shown a cartoon-style visual puzzle depicting a crime scene (see Figure S1). Instructions stated that a robber wearing blue pants, a striped hat, and a moustache had stolen a watch. Participants were asked to identify the correct suspect from five visually similar characters using a slider-based response interface (values 1–5). A countdown timer of two minutes was displayed at the top of the screen to induce time pressure. If no response was submitted before the timer expired, the system automatically recorded the default slider position.

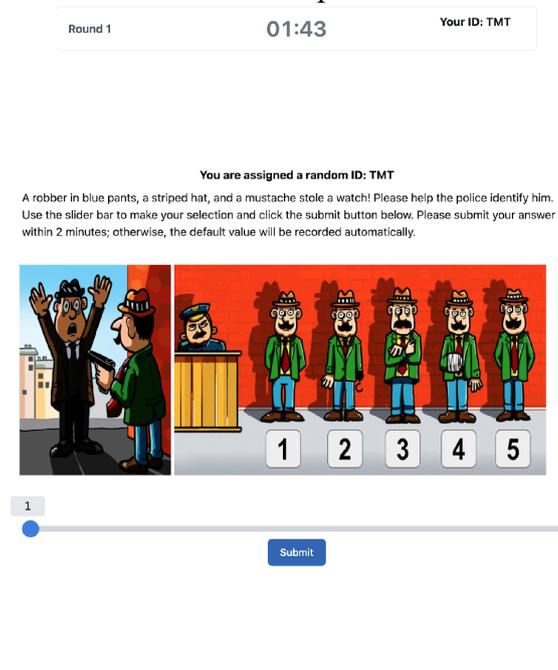

Figure S1. Round 1 Individual Robber Identification Task Interface.

This round served to familiarise participants with the task mechanics and to establish an individual performance context that would later justify managerial evaluation.

*S10.3 Transition to Round 2: Team Collaboration Instructions*

After completing Round 1, participants were informed that they would proceed to Round 2, which involved collaboration with two randomly assigned teammates who had also completed Round 1. Participants were instructed that all three team members should agree on a final answer and submit the same response. A chatbox interface appeared on the right side of the screen to facilitate discussion. Each participant was paired with two other players who had also completed Round 1, forming a three-person team for the collaborative task.

Participants were also informed that, following Round 2, the best-performing player would receive an additional monetary bonus (£0.50), increasing the motivational salience of the forthcoming managerial decision.

*S10.4 Manager Assignment and Team Task*

Before the collaborative task began, participants were informed that their three-person team would be assigned a manager. In both experiments reported here, participants were assigned an AI manager. Participants were told that the AI manager would evaluate team performance and select the best player based on performance in the task.

Participants then completed the same robber-identification task as in Round 1, now framed as a team activity. The visual stimulus and response interface were identical to those used previously (see Figure S2). Participants selected a suspect using the slider and submitted their guess. Participants were instructed to coordinate with their two teammates and submit a shared final answer; however, responses were submitted individually through the interface.

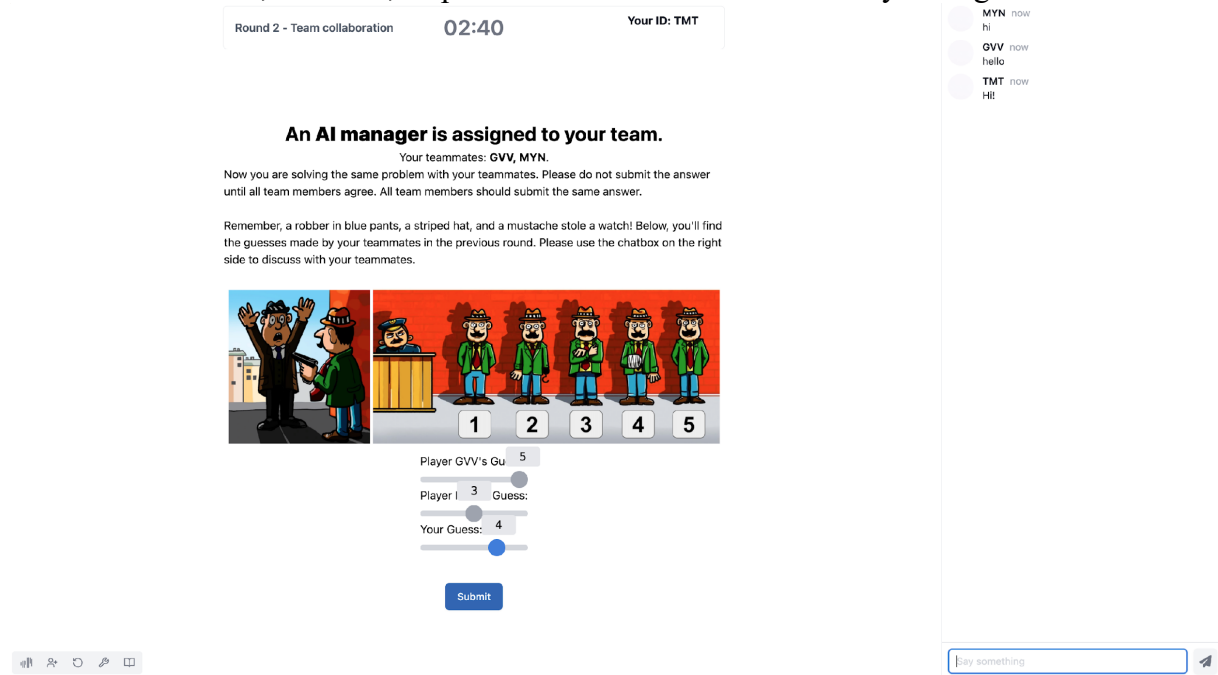

Figure S2. Round 2 Team Collaboration Task Interface With AI Manager Assigned.

*S10.5 Manager Evaluation Phase*

After submitting their Round 2 response, participants were informed that the AI manager would evaluate team performance and select the best player. Participants viewed a waiting screen indicating that the manager was making a decision. This delay was included to reinforce the perception of deliberative evaluation.

*S10.6 Final Decision Announcement*

Participants then saw a final decision screen stating whether they had been selected as the best player. If selected, the screen explicitly announced the participant's anonymised ID as the chosen player. In reality, selection was random, but participants were unaware of this during the experiment. This manipulation constituted the outcome valence factor (favourable vs. unfavourable decision) in both experiments.

*S10.7 Post-Task Evaluation*

Immediately after the decision, participants completed post-task surveys assessing their perceptions of the manager. All items used 11-point response scales (0–10) and were mandatory. The content of these evaluations differed slightly between experiments, as described below.

*S10.8 Experiment I (Text-Based AI Manager): Representation-Specific Details*

In Experiment I, the AI manager was represented exclusively through text. Prior to receiving the evaluation outcome, participants were shown a short textual description of the AI manager that manipulated perceived competence (high vs. low). Manager gender (female, male, or unspecified) was conveyed through pronouns only. No visual or anthropomorphic cues were provided. After the decision, participants evaluated the manager on trustworthiness, competence, fairness, leadership suitability, teammate suitability, and willingness to work with the manager. Participants also completed attention-check items identifying the manager's type (AI vs. human) and gender.

*S10.9 Experiment II (Image-Based AI Manager): Representation-Specific Details*

In Experiment II, the textual competence description was replaced with a facial image representing the AI manager (see Figures S3 and S4). These images were generated using a reverse-correlation paradigm and varied in perceived gender appearance and competence. No explicit gender labels were provided; instead, gender cues were conveyed implicitly through facial features. Following the decision, participants first rated the perceived femininity and masculinity of the manager's face and then evaluated the manager on the same outcome dimensions as in Experiment I. Apart from the representational modality and the inclusion of perceived gender ratings, all task procedures and interface elements were identical to Experiment I.

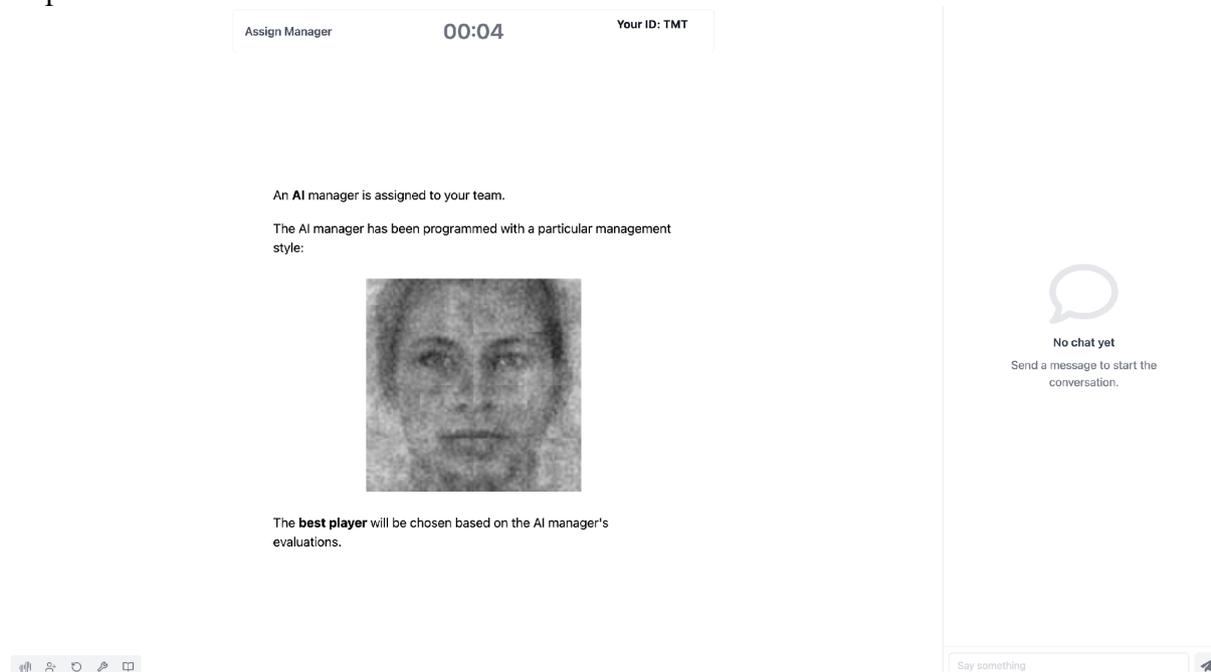

Figure S3. AI Manager Assignment Screen With Facial Representation Used in the Image-Based Condition.

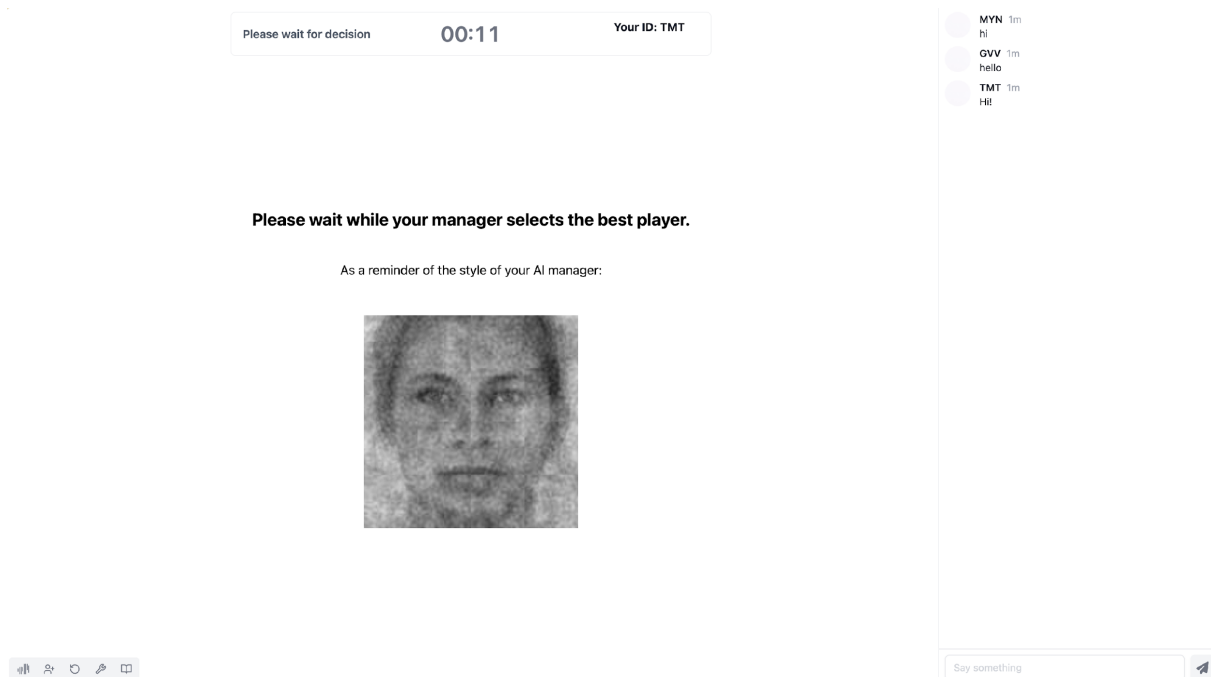

Figure S4. Manager Evaluation Waiting Screen With Visual Reminder.

*S10.10 Rationale for Unified Task Design*

The two experiments were designed to be procedurally equivalent so that any differences in evaluation outcomes could be attributed to the representational modality of the AI manager rather than to differences in task structure, incentives, or interaction flow. By holding the collaborative task, reward structure, timing, and evaluation sequence constant, the design isolates the causal role of textual versus anthropomorphic visual representation in activating competence-based versus gender-based evaluation heuristics.

**S11. Validation of Generated Faces.**

An exploratory factor analysis (EFA) using principal component analysis with varimax rotation was conducted to identify the underlying structure of nine interpersonal attributes: dominance, trustworthiness, attractiveness, warmth, fairness, leadership, willingness to work together, teamwork, and competence.

The data were suitable for factor analysis. The Kaiser-Meyer-Olkin (KMO) measure of sampling adequacy was 0.83, indicating adequate sampling (Kaiser, 1974; Shrestha, 2021). Bartlett's test of sphericity was significant, $\chi^2(36) = 1007.61$, $p < .001$, confirming that the correlation matrix was factorable.

Two factors with eigenvalues greater than 1.0 were extracted, together explaining 64.40% of the total variance. The first factor accounted for 42.59% of the variance, and the second accounted for 21.81% after rotation.

As shown in the rotated component matrix, trustworthiness (.77), warmth (.88), fairness (.83), willingness to work together (.59), teammate (.73), and attractiveness (.64) loaded strongly on the first factor, which was interpreted as Warmth. The second factor was defined by high loadings for dominance (.88), leadership (.63), and competence (.54), and was labelled Competence.

Communalities ranged from .44 (competence) to .79 (dominance and warmth), indicating that the extracted factors explained a substantial proportion of the variance across variables. Overall, the results suggest that perceptions of interpersonal attributes can be meaningfully summarised by two broad dimensions: Warmth and Competence.

Then we conducted an independent t-test to examine whether our reverse correlation paradigm successfully produced faces differing in femininity and competence across AI gender and competence conditions (Table 2a-2c). To aid interpretation of these results, a supplementary visualisation plots mean perceived competence against mean perceived femininity for each AI manager condition (see Figure S5). This figure provides a descriptive summary of the pattern reported in Tables 2a-2c and visually illustrates the separation of competence and gender cues across conditions.

### *S11.1. Feminine Appearance:*

Faces systematically varied in perceived femininity according to both AI gender and competence. Female faces were consistently rated as more feminine than neutral and male faces across both high- and low-competence conditions (see Table S1). Neutral faces were rated more feminine than male faces in the high-competence condition, but not significantly in the low-competence condition. Overall, these results indicate that the manipulation successfully produced gendered differences in perceived femininity, with the largest effects observed between female and male faces.

### *S11.2. Combined-Competence:*

Faces generated for high competence were rated as significantly more competent than those generated for low competence across all AI genders (see Table S2). The effect was strongest for male faces, followed by neutral and female faces. These results confirm that our competence manipulation was effective. In addition, low-competence faces were visually more blurred and pixelated than high-competence faces. This pattern likely reflects differences in the consistency and clarity of participants' underlying mental representations. High-competence individuals tend to elicit a more uniform and well-defined cognitive prototype, which enables reverse-correlation procedures to produce smoother, more coherent composite images. In contrast, low-competence individuals evoke more diffuse and valence-laden mental representations, with greater variability in the traits attributed to them across observers (Jaeger et al., 2024). This variability produces noisier, less stable selections during the image-generation phase, resulting in composites that appear less sharp. The visual differences between high- and low-competence composites therefore provide an additional implicit validation of the competence manipulation, as they reflect systematic differences in observers' shared mental templates for competent versus less competent social targets.

### *S11.3. Combined-Warmth:*

Warmth was measured as an exploratory variable because previous research suggests that perceived warmth and competence often co-occur and may influence each other (Fiske, 2018; Fiske et al., 2002; McKee et al., 2024). Although we did not manipulate warmth directly, we examined whether our competence manipulation affected warmth ratings (see Table S3). Results were mixed: only male faces showed a significant difference in warmth between high- and low-competence conditions, whereas female and neutral faces did not. This suggests that warmth was not reliably influenced by the competence manipulation, consistent with its exploratory role.

In sum, our manipulation successfully created faces that varied in femininity and competence, which were the primary variables of interest. Warmth was included as an exploratory measure to examine potential covariation with competence, but was not consistently affected by the manipulation.

After confirming that our visual stimuli successfully captured perceived differences in competence and gender, we tested how these faces affected participants' evaluations of AI managers in an interactive task that paralleled the text-based experiment.

**Table S1: Feminine Ratings by AI Competence Level and Gender**

| AI Competence | Gender Comparison | Mean | t | p | Cohen's d |
|---|---|---|---|---|---|
| High | Female vs Neutral | 7.08 vs 5.03 | 4.23 | **<.001** | 1.00 |
| High | Female vs Male | 7.08 vs 2.69 | 9.37 | **<.001** | 2.21 |
| High | Neutral vs Male | 5.03 vs 2.69 | 4.34 | **<.001** | 1.02 |
| Low | Female vs Neutral | 7.50 vs 6.11 | 2.91 | **.005** | 0.69 |
| Low | Female vs Male | 7.50 vs 5.16 | 4.54 | **<.001** | 1.06 |
| Low | Neutral vs Male | 6.11 vs 5.16 | 1.68 | .097 | 0.39 |

**Table S2: Combined-Competence Ratings by AI Gender**

| AI Gender | Competence Level | Mean | t | p | Cohen's d |
|---|---|---|---|---|---|
| Female | High vs Low | 6.61 vs 5.74 | 3.11 | **.003** | 0.73 |
| Neutral | High vs Low | 6.09 vs 5.12 | 3.35 | **.001** | 0.79 |
| Male | High vs Low | 6.18 vs 4.28 | 5.80 | **<.001** | 1.36 |

**Table S3: Combined-Warmth Ratings by AI Gender**

| AI Gender | Competence Level | Mean | t | p | Cohen's d |
|---|---|---|---|---|---|
| Female | High vs Low | 6.16 vs 6.39 | -0.81 | .418 | -0.19 |
| Neutral | High vs Low | 5.84 vs 5.37 | 1.51 | .136 | 0.36 |
| Male | High vs Low | 5.75 vs 4.73 | 2.74 | **.008** | 0.64 |

To complement the statistical analyses reported in Tables S1-S3, we provide a visual summary of the manipulation effects on perceived competence and femininity in Supplementary Figure S5. The figure plots mean perceived competence against mean perceived femininity for each AI manager condition, allowing a direct visual comparison across gender (female, male, neutral) and competence (high vs. low) levels.

Each point represents a condition-level mean, labelled using the following notation: F/M/N = Female/Male/Neutral appearance; H/L = High/Low competence; C = Competence. This visualisation is intended as a descriptive aid only and does not replace the inferential statistical tests reported in the main text and tables. As shown in Supplementary Figure S5, faces generated to represent high competence cluster at higher competence ratings across all genders, whereas femininity ratings vary systematically by gender condition, providing a visual confirmation of the effectiveness of the reverse-correlation manipulation. Warmth is not depicted in the visualization, consistent with its exploratory role in the analyses.

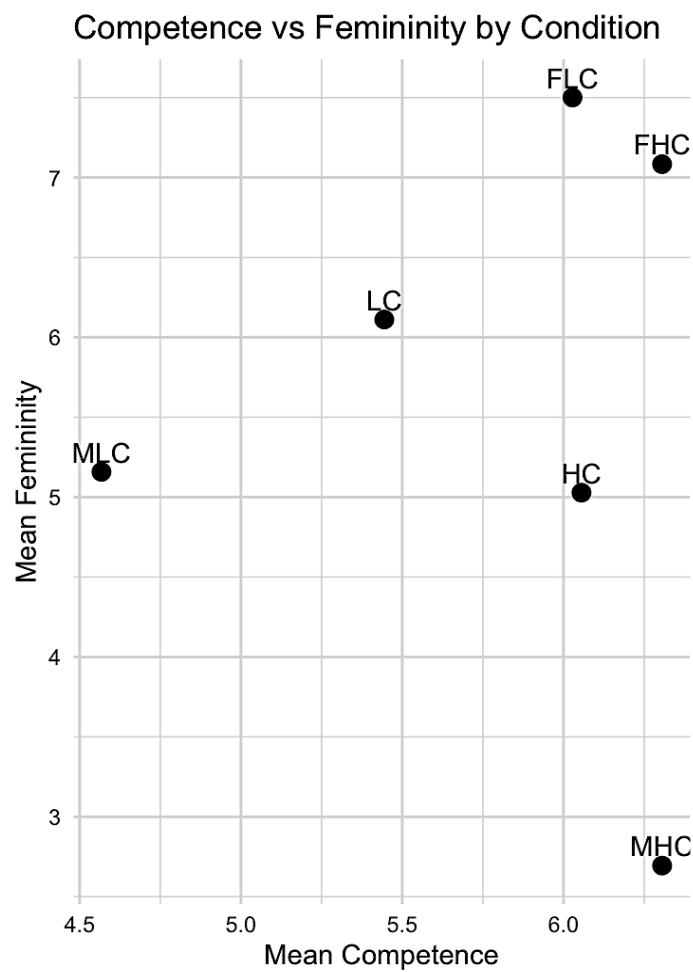

Figure S5. Mean Perceived Competence and Femininity of Generated AI Manager Faces by Condition.